\documentclass[10pt, conference, letterpaper]{IEEEtran}
\IEEEoverridecommandlockouts
\usepackage{cite}
\usepackage{amsmath,amssymb,amsfonts}
\usepackage[noend]{algorithmic}
\usepackage{graphicx}
\usepackage{textcomp}
\usepackage{xcolor}
\def\BibTeX{{\rm B\kern-.05em{\sc i\kern-.025em b}\kern-.08em
    T\kern-.1667em\lower.7ex\hbox{E}\kern-.125emX}}

\usepackage{mathrsfs}
\usepackage[ruled,linesnumbered,lined,boxed,commentsnumbered]{algorithm2e}
\usepackage{multirow}
\usepackage{booktabs}
\usepackage{bigstrut}
\usepackage[hang]{subfigure}
\usepackage{verbatim}

\usepackage{url}

\begin{document}

\title{MaaSO: SLO-aware Orchestration of Heterogeneous Model Instances for MaaS\\
}

\author{\IEEEauthorblockN{Xuan Mo}
\IEEEauthorblockA{\textit{Sun Yat-sen University} \\
\textit{School of Computer Science}\\
\textit{and Engineering}\\
GuangZhou, China}
\and
\IEEEauthorblockN{Yue Zhang}
\IEEEauthorblockA{\textit{Sun Yat-sen University} \\
\textit{School of Computer Science}\\
\textit{and Engineering}\\
GuangZhou, China}
\and
\IEEEauthorblockN{Weigang Wu}
\IEEEauthorblockA{\textit{Sun Yat-sen University} \\
\textit{School of Computer Science}\\
\textit{and Engineering}\\
GuangZhou, China}
}

\maketitle

\begin{abstract}

  Model-as-a-Service (MaaS) platforms face diverse Service Level Objective (SLO) requirements stemming from various large language model (LLM) applications, manifested in contextual complexity, first-token latency, and between-token latency. On the other hand, an LLM instance, when configured with different parallelism strategies and inference batch sizes, exhibits distinct performance characteristics and can thus be used to serve different SLO requirements. However, current LLM inference systems typically deploy instances of the same model with identical configurations, lacking mechanisms to leverage such heterogeneity. To fill this research gap, we propose MaaSO, the first ``MaaS Orchestrator'', which comprises three modules:  (1) a profiler characterizing instance performance under diverse parallelism strategies and inference batch sizes; (2) a placer optimizing heterogeneous instance configurations; (3) a distributor enabling SLO-aware request distribution and preventing cascaded timeouts in continuous batching. Experiments show that MaaSO improves the SLO satisfaction ratio by 15–30\% and reduces response latency by 40–60\% compared to existing approaches, and significantly lowers overall orchestration overhead. 
\end{abstract}


\section{Introduction}\label{sec:intro}

The proliferation of large language model (LLM)\cite{LLM_Survey} serving systems has catalyzed the emergence of Model-as-a-Service (MaaS)\cite{MaaS_Survey} platforms, where users access pre-trained models via cloud APIs without managing LLM post-training or instance deployment process. This paradigm shift significantly lowers the barrier for users to utilize state-of-the-art AI capabilities, driving their adoption across diverse LLM applications including intelligent agents\cite{agent_survey}, multimodal content generation\cite{agent_doc_survey}, and decision support systems\cite{agents_Industrial_1}. 

Such diverse applications inherently impose different SLO requirements\cite{DARKBIRD} on MaaS platforms. For example, intelligent agents in industrial process control\cite{agents_Industrial_3,agents_Industrial_2} require near-instant responses (e.g., sub-second latency for $<$ 50 tokens) for real-time decision making, while office document generators\cite{agents_doc_1,agents_doc_2} required handling massive, multi-modal input data to generate long-form content  (e.g., 2K-8K tokens with minutes-level deadlines). These variations manifest in two critical dimensions: (1) Contextual complexity, i.e., from short responses to lengthy content generation, and (2) Latency sensitivity, where inference request requirements vary significantly in terms of first-token and between-token latency\cite{MaaS_Survey}. Therefore, the diversity in SLO requirements significantly increases the complexity of MaaS management. 

Many efforts have been made to improve SLO satisfaction from different angles. AlpaServe\cite{Alpaserve} attempts to increase SLO satisfaction ratio by optimizing instance parallelism strategies. Meanwhile, Huang\cite{DARKBIRD} proposes to prioritize requests based on their SLO requirements and further determine execution order. However, all these studies consider homogeneous parallelism configurations for instances of the same model.  


On the other hand, the serving capability of a running instance is strongly influenced by two key factors: the parallelism strategy\cite{Alpa,Alpaserve} and the inference batch size\cite{DARKBIRD}. Intuitively, by configuring different instances of one model with distinct parallelism strategies and different inference batch sizes, we can more effectively serve MaaS workloads with diverse SLO requirements. However, due to the aforementioned contextual complexity and latency sensitivity, such heterogeneous instance orchestration faces three fundamental challenges: (1) How is the performance of an instance affected by heterogeneous configurations, under varying workload levels? (2) How to determine the optimal instance configuration and resource allocation  to better utilize system resources? (3) How to distribute diverse requests to heterogeneous instances to realize high SLO satisfaction ratio?  

To tackle these challenges, we propose MaaSO - the first MaaS orchestrator enabling efficient heterogeneous instance orchestration, which comprises three coordinated modules.  Our core contributions are summarized as follows:

\begin{enumerate}
    \item We establish a profiler for modeling the performance of inference instances under varying configuration factors (i.e., parallelism strategy and inference batch size), and different workload levels. We conduct real-world performance testings and then derive an instance throughput decay function.
    
    \item We design a placer with two coordinated algorithms to determine the optimal configurations for different instances: a simulator-guided algorithm that employs configuration-tree pruning and saturated-set mechanisms to reduce search complexity, and a dynamic programming algorithm that optimizes resource allocation across sub-clusters by maximizing service performance via serving score evaluation.
    
    \item We develop a distributor that enables efficient SLO-aware request distribution across heterogeneous instances while preventing cascaded timeouts in continuous batching. This modular separation allows complex instance configurations without compromising request distribution efficiency.
\end{enumerate}


\section{Background and Related works}
\subsection{Background}\label{sec:background}

Our work focuses on the management of heterogeneous model instances, where two core factors shape serving capabilities: parallelism strategies and inference batch sizes. Below, we introduce background knowledge about these aspects and their implications for LLM serving.

\textit{Instance Parallelism Strategy:} Several parallelism strategies have been proposed to enhance the utilization of accelerators (e.g., GPUs) in training and inference of deep learning models. Data Parallelism (DP) replicates models across multiple GPUs with partitioned input data\cite{DeepSpeed}, making it the most widely adopted strategy in deep learning. However, DP requires each GPU to store the entire model computation graph, thus necessitating combining it with other strategies for LLM serving. Tensor Parallelism (TP) and Pipeline Parallelism (PP), pioneered by NVIDIA's Megatron-LM\cite{megatron-LM}, distribute individual models across devices by partitioning computation graphs within or across layers. While GPipe\cite{GPipe} mitigates pipeline bubbles through micro-batching, Pipedream\cite{Pipedream} further optimizes backward scheduling, with both techniques primarily targeting PP in LLM training. LLM inference engines like DeepSpeed\cite{DeepSpeed}, vLLM\cite{vLLM}, and SGLang\cite{SGLang} integrate all the three strategies. vLLM implements PagedAttention and Continuous Batching\cite{Orca}, while SGLang reduces request response latency through shared prefix caching (via a radix-tree) during the prefill phase. These engines set the current standard for inference.

\textit{Inference Batch Size:} To ensure reasonable completion times for user requests, it is impractical for an instance to process all continuously batched requests concurrently without constraints. Excessive workload level can cause severe GPU resource contention (e.g., memory bandwidth saturation or compute unit overload\cite{DynaServe}), leading to unpredictable decoding latency spikes. LLM inference engines usually enforce basic performance isolation by artificially limiting the maximum batch size (i.e., the upper bound of concurrently processed requests) of the request queue that an instance can handle. Specifically, vLLM implements this via the parameter \texttt{max-num-seqs}, while SGLang employs \texttt{max-running-requests} for a comparable mechanism. When this processing queue reaches its capacity, newly arriving requests are redirected to a waiting queue.

While it is easy to determine the parallelism strategy and inference batch size for the performance of single instance, current inference engines generally lack native multi-instance orchestration capabilities at the cluster level. Efficient operation of LLM inference system still relies on external coordination middleware (e.g., Kubernetes\cite{Kubernetes} or NVIDIA Dynamo\cite{dynamo}) and system managers for workload distribution and instance management, incurring operational overhead for systematic orchestration.  Consequently, implementing adaptive parallel model deployment services on cloud clusters and optimizing performance based on user-perceived quality of service (QoS) have gradually gained attention among researchers in LLM inference systems.

\subsection{Related Works} \label{sec:related}
Several approaches have been proposed to automate the selection of parallelism strategies and deployment of LLMs. Megatron-LM v2\cite{Megatron-LM-v2} combines inter-op and intra-op parallelism through manually specified strategies to support distributed training of a 530B LLM. Microsoft's DeepSpeed\cite{DeepSpeed} introduces ZeRO technology, which optimizes GPU memory usage in LLM training and inference through multi-level data offloading strategies. FlexFlow\cite{FlexFLow} proposes a novel execution simulator to predict the performance of model instances under different parallelism strategies and identify optimal deep learning (DL) model deployment configurations. Alpa\cite{Alpa} leverages cluster device characteristics to partition LLM into segments, using dynamic programming for PP stage mapping, and employs Integer Linear Programming (ILP) for TP degree decisions. While these algorithms enable automated parallelism strategy selection for large deep learning models, they do not consider multi-instance scenarios nor do they support SLO-aware model instance configurations. 

AlpaServe\cite{Alpaserve} investigates how model parallelism strategies affect request SLO satisfaction in multi-model inference services. However, it assumes homogeneous parallelism for a single model and does not consider diverse SLO requirements for the same model.

\begin{figure*}[!ht]
    \centering
    \includegraphics[width=0.9\linewidth]{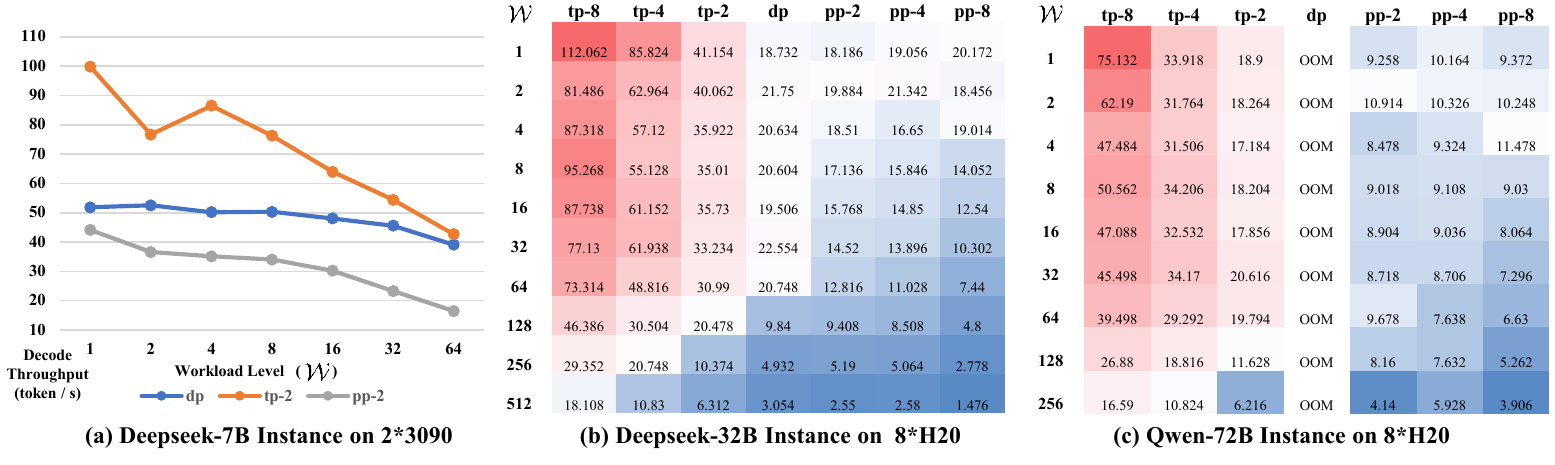}
    \includegraphics[width=0.9\linewidth]{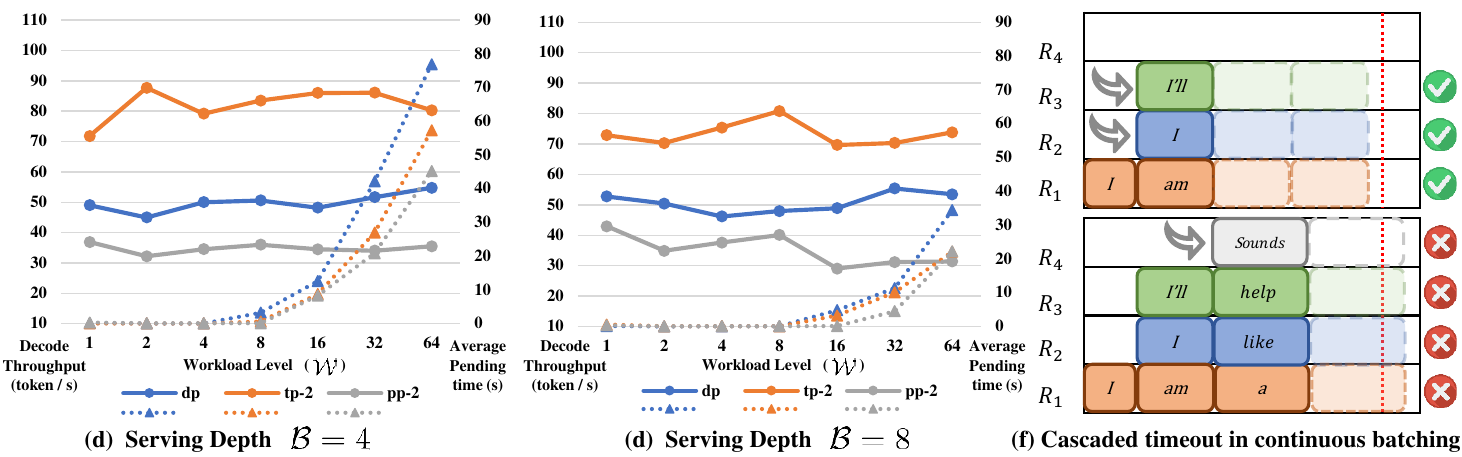}
    \caption{The decoding throughput of a serving LLM instance is determined by its parallelism strategy and the workload level (denoted as $\mathcal{W}$, the real-time number of concurrent requests). By tuning the inference batch size (denoted as $\mathcal{B}$, a configurable parameter of the instance), an instance can trade off its decoding throughput against its response latency.}
    \label{fig:Instance_interference}
\end{figure*}

    


\section{Performance Analysis of LLM Configuration} \label{sec:sys_analysis}

By conducting experiments, we analyze how different parallelism strategies and inference batch size settings impact instance serving performance under varying workload levels. These experiments are conducted using the vLLM inference engine on GPU servers detailed in Section~\ref{sec:exp_setting}.

\subsection{Impact of Instance Parallelism Strategy}
Fig.~\ref{fig:Instance_interference} presents the decoding throughput measurements across different parallelism strategies for three representative LLM models: DeepSeek-7B, DeepSeek-32B\cite{DeepSeek}, and Qwen-72B\cite{Qwen}. We specify parallelism strategies using the notation (strategy type)-(degree), e.g., ``tp-2" for tensor parallelism with degree 2, and ``dp" for data parallelism. 



Our analysis reveals two critical patterns:
\begin{itemize}
    \item \textbf{Throughput Degradation}: Although fluctuations due to the stochastic nature of the LLM's output length exist, the throughput of parallel instances exhibits a logarithmic decay as workload level $\mathcal{W}$ increase. This degradation is particularly pronounced in high-degree parallelism strategies (e.g., tp-8, pp-8).
    
    \item \textbf{Performance Convergence}: When workloads are saturated, the performance of higher-degree parallel instances may degrade to a level comparable to that of lower-degree parallel instances. For example, in sub-diagonal regions of Fig.~\ref{fig:Instance_interference}-(b) and (c), a tp-8 instance processing 512 concurrent requests achieves similar throughput to multiple tp-4 or tp-2 instances handling the same workload, with load-balanced workload level of 256 and 128, respectively.
\end{itemize}

\subsection{Impact of Inference Batch size} \label{sec:latency_vs_depth}

While maximizing global throughput suggests serving more concurrent requests with fewer instances, we observed that the batch size of each instance must be carefully controlled so as to prevent cascaded timeout failures, a critical phenomenon in continuous batching systems. As shown in Fig.~\ref{fig:Instance_interference}-(f), when requests ($R_1, R_2, R_3$) are initially processed on schedule, the admission of a new request $R_4$ can delay the entire batch due to resource contention, causing all requests to timeout. Since the admission of new requests gradually reduces the decoding speed of early arriving requests, the unpredictability of real-time workload levels $\mathcal{W}$ makes SLO guarantees for inference requests particularly difficult.

\begin{figure}[htbp]
    \centering
    \includegraphics[width=0.8\linewidth]{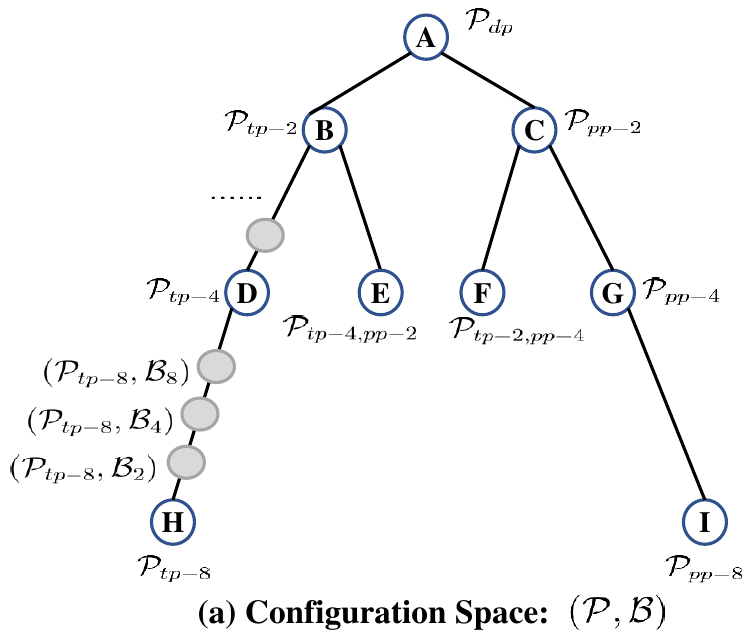}
    \includegraphics[width=0.85\linewidth]{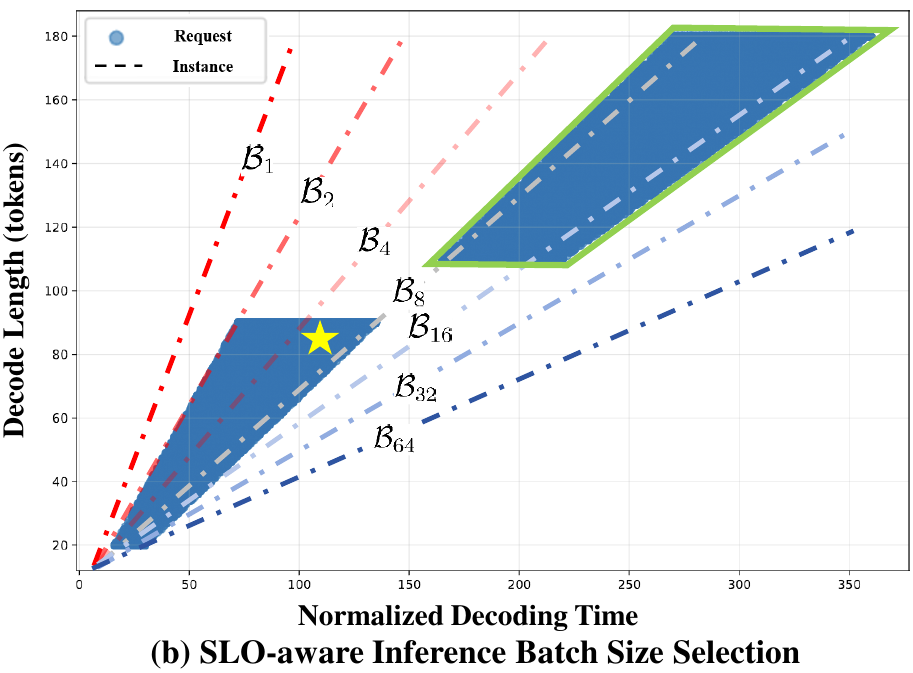}
    \caption{Optimized search process for instance configuration} \label{fig:ins_deploy_space}
\end{figure}

As described in Section \ref{sec:background}, the inference batch size $\mathcal{B}$ serves as a key configuration for mitigating throughput degradation under concurrent workloads. Fig.~\ref{fig:ins_deploy_space}-(d) and Fig.~\ref{fig:ins_deploy_space}-(e) compare two Deepseek-7B instances with different inference batch sizes (set via vLLM's API):

\begin{itemize}
    \item With lower batch size ($\mathcal{B}=4 $ or $ \mathcal{B}_4$), the instance achieves higher per-request decoding throughput. However, as concurrent requests accumulate, the queuing delay increases dramatically.
    \item Increasing batch size to $ \mathcal{B} = 8$ significantly mitigates the queuing delay without substantial throughput degradation, demonstrating the delicate balance between latency and throughput.
\end{itemize}

The above analysis reveals that inference batch size, like parallelism strategy, is a critical configuration parameter that requires carefully tuning. The optimal batch size of an instance must account for both the instance's parallelism strategy and the orchestrator's workload distribution strategy, to prevent either excessive queuing or inefficient resource utilization.

\subsection{Instance Configuration v.s. Request SLO}

In Fig.~\ref{fig:ins_deploy_space}-(a), we organize the instance configuration parameters $(\mathcal{P}, \mathcal{B})$ into a configuration tree. Each blue node represents a parallelism strategy $\mathcal{P}$, while gray nodes indicate different batch sizes $\mathcal{B}$ under that strategy. Through in-order traversal of this tree, we obtain deployment configurations in decreasing order of decoding throughput.

To analyze the relationship between instance configurations and request SLO requirements, we establish a two-dimensional framework as shown in Fig.~\ref{fig:ins_deploy_space}-(b). For a request $r$, the x-axis represents the  normalized deadline, which is calculated as $\tau_r = d_r \cdot \theta_r \cdot \theta$, where $d_r$ is the decoding length, $\theta_r$ is the request's SLO factor, which quantifies its deadline urgency, $\theta$ is the time slice corresponding to the single-token decoding latency of a DP-instance$(\mathcal{P}_{dp}, \mathcal{B}_1)$. This deadline setting follows the normalized deadline scheme\cite{Orca,DARKBIRD}. Each instance's decoding performance boundary (dashed line) is modeled as $y = T_i(x - p_r)$, where $T_i$ represents the decoding throughput of instance $i$ when its inference batch size saturated (i.e., $\mathcal{W}_i = \mathcal{B}_i$), and $p_r$ captures request pending time.

Obviously, only requests to the right of each boundary can be completed by their corresponding instances without triggering cascaded timeouts. When increasing the inference batch size from $\mathcal{B}_4$ to $\mathcal{B}_8$, the instance's throughput degradation creates a request rejection zone (with the yellow pentagram). While higher inference batch sizes reduce the response latency of concurrent requests, they also reduce effective throughput. When request SLOs are similar (e.g., requests in green boxes), selecting between $\mathcal{B}_4$ and $\mathcal{B}_8$ appears straightforward. However, when diverse request clusters coexist, the configuration selection becomes complex, and deploying many $\mathcal{B}_2$ replicas would cause significant resource waste.

The above analysis reveals that deploying heterogeneous instances can better serve diverse SLO requirements than homogeneous configurations. Strategic deployment of specialized instances with different $(\mathcal{P}, \mathcal{B})$ configurations can create composite service regions capable of satisfying the full spectrum of SLO demands, providing the foundational motivation for our heterogeneous instance orchestration approach.

\section{MaaSO Design}

\subsection{Overview}\label{sec:system_overview}




Inspired by the analysis in the previous section, we design MaaSO to efficiently orchestrate heterogeneous model instances for better serving inference requests with diverse SLOs. 

In this study, we assume that each LLM instance is deployed on exclusive GPU sets. Given that LLM inference engines are engineered to fully saturate assigned accelerator resources, co-deployment of multiple instances may lead to inefficiency.  



\begin{figure}[!ht]
    \centering
    \includegraphics[width=1.0\linewidth]{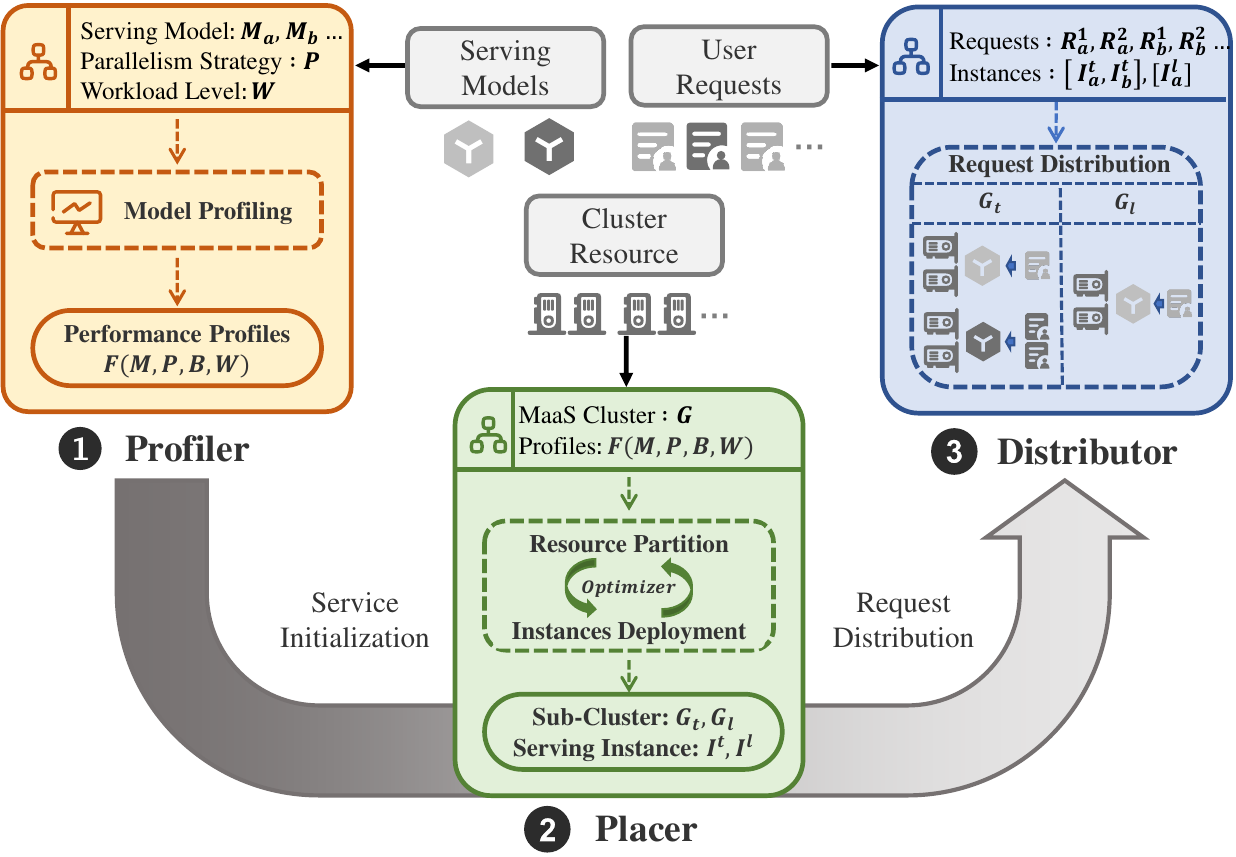}
    \caption{Example workflow of MaaSO with two sub-clusters.}
    \label{fig:System_overview}
\end{figure}

As shown in Fig.~\ref{fig:System_overview}, MaaSO consists of three main modules: the profiler, the placer, and the distributor. The profiler quantifies instance serving capability under varying parallelism strategies and different inference batch sizes. The placer partitions GPU sub-clusters and deploys heterogeneous instances across sub-clusters. 
Such partitioning is for the purpose of simplifying the complexity of instance management. Our approach can also handle instance heterogeneity at the granularity of individual servers. The distributor efficiently distributes requests to sub-clusters based on their SLO requirements and further maps them to appropriate instances. 

In the following, we describe the details of MaaSO. The three modules are introduced one by one.  Since the placer is more complex than the other two modules, we use three sub-sections to describe the optimizer formulation, configuration scoring and configuration selection, respectively.






\subsection{Profiler: Performance Modeling}\label{sec:Profiler}

As discussed in Section \ref{sec:sys_analysis}, instance configuration and workload level may significantly influence the decoding performance of an instance. Therefore, the profiler must evaluate the serving capability of an instance (across different models $M \in [M_a, M_b, ...]$) under various instance configurations $(\mathcal{P}, \mathcal{B}) \in ([\mathcal{P}_{dp}, ...] \times [1,2, ...]) $ and workload levels $W \in [1,2, ...]$. However, exhaustive performance profiling across this parameter space would trigger a massive number of parallel instance launches, making the process prohibitively expensive. Therefore, we derive throughput decay functions from the performance analysis in Section \ref{sec:sys_analysis}, so as to avoid significant profiling overhead.

Based on the results in Fig.~\ref{fig:Instance_interference}, we make two key insights. First, the decoding throughput of an instance exhibits different decay patterns with different models or parallelism strategies as workload level accumulates. Second, the inference batch size $\mathcal{B}_i$ of an instance $i$ will truncate its logarithmic throughput decay when workload level $\mathcal{W}_i$ reaches $\mathcal{B}_i$. So, we propose the throughput decay function:
\begin{equation}\label{eq:throughput_model}
  F(M,\mathcal{P},\mathcal{B},\mathcal{W})=T_0(M,\mathcal{P}) \cdot (1 - \delta_{\mathcal{P}} \cdot \log(\epsilon_{\mathcal{P}} + \min(\mathcal{B},\mathcal{W}))) ,
\end{equation} 
where $T_0(M,\mathcal{P})$ represents the baseline throughput of an instance of model $M$ that serves one request under the parallelism strategy $\mathcal{P}$. The parameters $\delta_\mathcal{P}$ and $\epsilon_\mathcal{P}$ quantify the sensitivity of the instance configured as $(M,\mathcal{P})$ to workload level. To derive this throughput decay function, the profiler employs a limited set of $(M,\mathcal{P},\mathcal{B},\mathcal{W})$ for throughput testing, allowing it to fit Eq.~\eqref{eq:throughput_model} and determine reliable values for $\delta_\mathcal{P}$ and $\epsilon_\mathcal{P}$ through least squares fitting.

\subsection{Placer: Optimizer Formulation}\label{sec:optimizer}

The optimizer manages three prioritized objectives based on user tolerance for request SLO violations. First, SLO satisfaction is assigned the highest priority to maximize request SLO attainment. Second, the system should minimize the response latency to avoid user abandonment. Third, optimizing the decoding throughput aligns with user preferences for efficient inference. To achieve these objectives, we leverage the throughput estimation function $F(\cdot)$ to model the end-to-end latency of an inference request $r$. 

Consider request $r \in R_a$ assigned to an instance $i \in I_a$; it must complete decoding $S_r$ tokens before deadline $\tau_r$. The end-to-end latency $L_r$ comprises queuing latency $L_q^{r}$ and decoding latency $L_d^{r}$:
\begin{alignat}{2} \label{eq:latency_component}
  L_r &= L^r_q + L^r_d , \\
  L_d^{r} &= \frac{\mathbb{E}_{r \in R_a} [S_r]}{F(M_a,\mathcal{P}_{i}, \mathcal{B}_i ,\mathcal{W}_{i})} , \quad L_q^{r} \propto \frac{\mathcal{W}^a \cdot \mathbb{E}_{r \in R_a} [L_d^r]}{\sum_{i \in I_a} \mathcal{B}_i  } ,\notag
\end{alignat}
where $\mathcal{W}_{i}$ represents the number of concurrent requests processed on instance $i$ when request $r$ arrives. Queuing occurs when $\mathcal{W}_i$ exceeds the inference batch size $\mathcal{B}_i$, at which point the throughput decay of $i$ is truncated. The decoding latency $L_d^r$ can be estimated by the throughput decay function $F$ when $\mathcal{W}_i$ is determined. Queuing latency $L_q^r$ is mainly affected by: the total workload level $\mathcal{W}^a$ of requests for model $a$, the expected decoding latency  $\mathbb{E}(L_d^r)$ of requests $r \in R_a$, and the inverse of the total inference batch size $\sum \mathcal{B}_i$ of all instances $i \in I_a$. SLO violation occurs when $L_r$ is greater than $\tau_r$. 

So, the optimizer manages three objectives: SLO satisfaction $\Phi_S$, system decoding throughput $\Phi_T$, and queuing latency $\Phi_L$:
\begin{alignat}{2}
\Phi_S : & \quad \max  \sum_{r \in R}  \sum_{i \in I} x_{r,i} \cdot \mathbb{I}\left[ L_q^{r} + L_d^{r} \leq \tau_r \right] \label{eq:obj_slo} ,\\
\Phi_T : & \quad \max\sum_{i \in I} \sum_{g \in G} y_{i,g} \cdot F(M_i, \mathcal{P}_i, \mathcal{B}_i, \mathcal{W}_i) \label{eq:obj_throughput} ,\\
\Phi_L : & \quad \min \sum_{r \in R} \sum_{i \in I} x_{r,i} \cdot \frac{\mathcal{W}_i \cdot \mathbb{E}[L^r_d ]}{\mathcal{B}_i} \label{eq:obj_latency} ,\\
\mbox{s.t.} \quad
& x_{r,i}, y_{i,g} \in [0,1]; \quad \mathcal{P}_i \in \mathcal{P}; \quad \mathcal{B}_i \in \mathbb{Z}^{+}; & \tag{a} \\
& \sum_{r \in R} x_{r,i} \leq 1 , \forall i \in I; \quad \sum_{i \in I} y_{i,g} \leq 1 , \forall g \in G; & \tag{b} \\
& \sum_{i \in I} y_{i,g} \cdot N(\mathcal{P}_i)\leq N(G); & \tag{c} \\
& \sum_{i \in I} y_{i,g} \cdot Mem(M_i, \mathcal{P}_i)\leq Mem(g) ,\forall g \in G. & \tag{d} 
\end{alignat}

Constraint (a) defines the binary decision variables $x_{r,i}$ and $y_{i,g}$, which determine the assignment of request $r$ to instance $i$ and the placement of instance $i$ on GPU $g$, respectively. $\mathcal{P}_i$ and $\mathcal{B}_i$ denote the parallelism strategy and inference batch size configured for each instance $i$. Constraint (b) contains two critical provisions: one that prevents requests from being duplicated in processing, and the other that guarantees GPU-level exclusivity in instance configuration. Constraint (c) maintains cluster-wide resource boundaries by ensuring that the total GPUs allocated to parallel instances ($\sum_i N(\mathcal{P}_i)$) do not exceed available resources ($|G|$). Constraint (d) enforces GPU memory limits by verifying the memory usage of each instance configuration decision ($\sum_i\text{Mem}(M_i, \mathcal{P}_i)$).

Directly solving this optimization problem is impractical for two fundamental reasons. First, the combinatorial complexity scales exponentially with problem size. Second, critical variables in Eq.~\eqref{eq:latency_component} cannot be reliably estimated - including real-time fluctuating workload level $\mathcal{W}_i$ and unpredictable LLM decoding lengths $S_r$. These dynamic unknowns make it impossible to model the above optimization objectives.

Instead, inspired by AlpaServe\cite{Alpaserve},  we adopt a simulator-based approach. The simulator models workload distribution and instance execution, checking if a request $r$ can meet $\tau_r$ (via Eq.~\eqref{eq:obj_slo}). This approach avoids intractable optimization while capturing core SLO constraints.

\subsection{Placer: Configuration Scoring}\label{sec:score_and_pruning}


While the simulator enables us to quantitatively evaluate the three optimization objectives by simulating the execution of requests, two critical issues still remain: (1) effectively balancing multiple optimization objectives, and  (2) efficiently defining candidate configurations $(\mathcal{P}, \mathcal{B})$ to reduce search complexity. To address these issues, we propose a quantitative serving score evaluation mechanism and two strategic pruning techniques. 
  
First, to unify these multiple simulation-observed metrics into a cohesive decision framework, we introduce a composite serving score $\Phi(I, R)$ that weights metrics via normalized evaluation functions:
\begin{align}\label{eq:score_simulation}
\Phi(I, R) &= \alpha \cdot \Phi^n_S (I, R) + \beta \cdot \Phi^n_T(I, R, \gamma_T) \nonumber \\
& \quad +  (1-\beta) \cdot \Phi^n_L(I, R, \gamma_L) , \\
 \Phi^n_T(I, R, \gamma_T) &= \frac{\min (\Phi_T(I, R), \gamma_T)}{\gamma_T} , \\
  \Phi^n_L(I, R, \gamma_L) &= \frac{\max (\gamma_L - \Phi_L(I, R), 0) }{\gamma_L} .
\end{align}

Here, $\Phi^n_{S}(\cdot)$, $\Phi^n_{T}(\cdot)$ and $\Phi^n_{L}(\cdot)$ represent the SLO satisfaction ratio of requests, normalized decoding throughput and response latency, respectively. $\gamma_T$ and $\gamma_L$ represent throughput and latency thresholds that normalize the original metrics observed from the simulator ($\Phi_T(I, R)$, and $\Phi_L(I, R)$) to the range $[0,1]$. They are set based on the maximum throughput achievable by parallel instances and the maximum acceptable latency. The weight $\alpha$ prioritizes SLO satisfaction (typically $\alpha > 2$), while $\beta \in [0,1]$ balances the preference between throughput and latency optimization. This parameterization enables MaaSO to align its configuration strategy with specific service priorities.

Second, for the configuration tree detailed in Fig.~\ref{fig:ins_deploy_space}-(a), we propose two pruning strategies based on observations regarding instance decoding performance:


\textit{Instance Parallelism Strategy Pruning}: By traversing the configuration tree in-order, we prune nodes whose configurations fail to outperform $\mathcal{P}_{dp}$ while consuming more resources. Based on the analysis of Fig.~\ref{fig:Instance_interference}-(a), we could terminate the configuration search at node $A$, since PP-instances cannot exceed DP-instance decoding throughput despite using more GPUs. For nodes $E , F$, evaluations show that they cannot match $\mathcal{P}_{tp-8}$ (node $H$), so we only evaluate TP/DP configurations when distributed configurations across servers are not adopted.

\textit{Inference Batch Size Pruning}: For each parallelism configuration, we identify and retain only inference batch sizes that create Pareto-optimal trade-offs between latency and SLO satisfaction. This is illustrated by the process of configuring instances serving the green boxed requests in Fig.~\ref{fig:ins_deploy_space}-(b): during configuration, we first filter out unnecessarily low batch sizes (e.g., $\mathcal{B}_1$ and $\mathcal{B}_2$) - these increase response latency without improving SLO satisfaction. We then eliminate excessively high batch sizes (e.g., $\mathcal{B}_{32}$ and $\mathcal{B}_{64}$), as they fail to meet any request's SLO under high workload level $\mathcal{W}$.



\begin{algorithm}[ht]
\caption{Simulator-Based Instance Configuration}
\begin{algorithmic}[1]\label{alg:ins_deployment}
\STATE \textbf{Input:} Request set $R_x$,  sub-cluster $G_x$, Model set $M$, Model profiles $F(\cdot)$.
\STATE \textbf{Output:} Optimal instance list $\mathcal{I}^*$, Score list $\Phi^*$.

\STATE  $\mathcal{I}^* \leftarrow [\emptyset] \times |G_x|$; $\Phi^* \leftarrow [0] \times |G_x|$;
\STATE $configs \leftarrow getPrunedConfigs(M,R_x, F)$;

\FOR{$(\mathcal{P}_i, \mathcal{B}_i) \in configs$}
    \STATE \textit{// Initialize configuration solution, saturated model}
    
    \textit{set, unserved request set and serving score.}
    \STATE $\mathcal{I} \leftarrow \emptyset ; M_{sat} \leftarrow \emptyset; R_{res} \leftarrow R_x; \phi \leftarrow 0$; 
    
    \WHILE{$gpuCount(\mathcal{I}) < |G_x|$ \textbf{and} $M_{sat} \neq M$}
        \STATE Update $R_{res} \leftarrow unservedRequests(\mathcal{I}, R_x$)\;
        \STATE $m' \leftarrow mostUnservedModel(R_{res}, M \setminus M_{sat}$)\;
        \STATE $\mathcal{I}' \leftarrow \mathcal{I}.addInstance(m', \mathcal{P}_i, \mathcal{B}_i)$\;
        \STATE $k \leftarrow gpuCount(\mathcal{I})$ \;
        \STATE $\phi' \leftarrow$ evaluating $\Phi (\mathcal{I}', R_x)$ with Eq.~\eqref{eq:score_simulation}\;
        \IF{$\phi' > \phi$}
            \STATE $\phi \leftarrow \phi'$; $\mathcal{I} \leftarrow \mathcal{I}'$; \textit{// Accept new instance}
        \ELSE
            \STATE $M_{sat}.addModel(m')$; \textit{// Stop deploying } $m'$
        \ENDIF

        \IF{$\phi > \Phi^*[k]$}
            \STATE $\Phi^*[k] \leftarrow \phi$; ~$\mathcal{I}^*[k] \leftarrow \mathcal{I}$;
        \ENDIF
    \ENDWHILE
\ENDFOR
\RETURN $\mathcal{I}^*, \Phi^*$
\end{algorithmic}
\end{algorithm}

\begin{algorithm}[htbp]
\caption{Dynamic Resource Partition}
\begin{algorithmic}[1]\label{alg:res_partition}
\STATE \textbf{Input:} GPU cluster $G$, Request set $R$, Model set $M$, Model profiles $F(\cdot)$.
\STATE \textbf{Output:} Sub-cluster $[G_t, G_l]$, Instance $[\mathcal{I}^t, \mathcal{I}^l]$
\STATE $R_t, R_l \leftarrow byRequestSLO(R)$;

\STATE \textit{// Initial resource allocation by request counts}
\STATE $ratio \leftarrow |R_l| / |R|$ ; $G'_{l} \leftarrow \lfloor ratio \times |G| \rfloor$;

\STATE \textit{// Deploy instances for both sub-cluster using Alg.~\ref{alg:ins_deployment}}
\STATE $\mathcal{I}_l^*, \Phi_l^* \leftarrow insDeploy(R_l, G'_l$); 
\STATE $\mathcal{I}_t^*, \Phi_t^* \leftarrow insDeploy(R_t, G)$

\STATE \textit{// System serving score initial}
\STATE $\Phi_{opt} \leftarrow \max (insDeploy(R, G))$; \textit{// if needed, or 0}
\FOR{$g_l = 1$ to $|G'_l|$}
    \STATE $g _t \leftarrow |G| - g_l$ ; $\phi_t \leftarrow \Phi_t^*[g_t]$ ; $\phi_l \leftarrow \Phi_l^*[g_l]$
        
    \IF{$\phi_t + \phi_l > \Phi_{opt}$}
        \STATE $\Phi_{opt} \leftarrow \phi_t + \phi_l$; 
        \STATE $G_t, G_l \leftarrow updatePartion(g_t, g_l)$;
        \STATE $\mathcal{I}^t, \mathcal{I}^l \leftarrow \mathcal{I}_t^*[g_t] , \mathcal{I}_l^*[g_l]$
    \ENDIF
\ENDFOR
\RETURN $[G_t, G_l]$, $[\mathcal{I}^t, \mathcal{I}^l]$
\end{algorithmic}
\end{algorithm}

\subsection{Placer: Configuration Selection}\label{sec:deploy_partiton}
MaaSO partitions the GPU cluster into specialized sub-clusters and determines optimal instance configurations for each sub-cluster, as illustrated in Fig.~\ref{fig:System_overview}. The placer employs two algorithms to realize such optimality.  Alg.~\ref{alg:ins_deployment} performs simulator-guided instance deployment for an individual GPU cluster, while Alg.~\ref{alg:res_partition} implements dynamic programming-based resource partitioning and coordinates configurations across sub-clusters, optimizing the overall service performance based on global serving score. 

We introduce the two algorithms using a two-sub-cluster example: one serving throughput-oriented requests $R_t$ with strict deadlines, and another serving latency-tolerant requests $R_l$ with relaxed SLO constraints (e.g., green boxed requests in Fig.~\ref{fig:ins_deploy_space}-(b)). This example of sub-cluster partitioning is sufficient to demonstrate workload distribution for diverse request SLOs. When more sub-clusters are needed, we can scale to orchestrate multiple sub-clusters via the same dynamic programming approach in Alg.~\ref{alg:res_partition}.

Alg.~\ref{alg:ins_deployment} iteratively configures model instances for any sub-cluster $G_x$ using a greedy heuristic. After selecting candidate configurations through pruning strategies (line 4), the algorithm employs a nested loop structure to evaluate each configuration. For each configuration $(\mathcal{P}_i, \mathcal{B}_i)$, the placer identifies the model $m'$ with the largest number of unserved requests from the remaining candidate set $M \setminus M_{sat}$ and creates a new deployment $\mathcal{I}'$ by adding an instance of $m'$ (lines 9-11). If the new instance improves the serving score $\phi$, it is accepted; otherwise, model $m'$ is added to the saturated set $M_{sat}$ to prevent further unproductive exploration (lines 13-16). The algorithm maintains optimal solutions $\mathcal{I}^*[k]$ and corresponding serving scores $\Phi^*[k]$ for all possible GPU allocations $k$ across all evaluated configurations (lines 17-19).

Alg.~\ref{alg:res_partition} coordinates sub-cluster resource allocation by partitioning the GPU cluster $G$ into specialized sub-clusters $[G_t, G_l]$ and determining optimal instances configurations $[I^t, I^l]$ for each sub-cluster. The algorithm first categorizes requests by SLO requirements (line 3), separating throughput-oriented requests $R_t$ from latency-tolerant requests $R_l$. Since serving requests with stricter deadlines typically requires higher-degree tensor parallelism or reduced inference batch size, the algorithm employs an asymmetric resource allocation strategy: it provides initial estimation only for sub-cluster $G_l$ based on request proportions (line 5), while allowing $G_t$ access to all GPUs during solution generation (line 8).

Then, Alg.~\ref{alg:ins_deployment} is invoked for both sub-clusters to obtain solution arrays with optimal configurations. The dynamic programming phase (lines 11-16) explores all feasible resource partitions $(g_t, g_l)$ where $g_t + g_l = |G|$, accessing pre-computed serving scores $(\phi_t, \phi_l)$ and maximizing their summation to determine the optimal partition. To ensure robustness, the algorithm initializes $\Phi_{opt}$ with a homogeneous deployment baseline (line 10), allowing it to revert to homogeneous strategies when heterogeneous deployment provides no performance gains. This would bring extra orchestration overhead, while simple zero initialization would be sufficient in most cases. In extreme cases, heterogeneous instance scheme may not yield performance gains. We will discuss these boundary conditions and their impact in the limitations analysis of Section \ref{sec:exp_analysis}.

The decomposition of these two algorithms offers significant computational advantages over joint optimization approaches. The overall complexity is dominated by Alg.~\ref{alg:ins_deployment}: The pruning strategies reduce the search space from $\mathcal{O}(|\mathcal{P}| \times |\mathcal{B}|)$ to approximately $\mathcal{O}(|\mathcal{P}| \times |\mathcal{B}|_{\text{valid}}/2)$, where the ratio of $|\mathcal{B}|_{\text{valid}}$ to $|\mathcal{B}|$ depends on the difference in SLO demand between the full request set $R$ and sub-set $R_x$. The dynamic programming adds only $O(|G|)$ operations,  to enable complete resource allocation space exploration. By pre-computing all possible deployment solutions, we avoid the exponential complexity of joint optimization.


\subsection{Distributor: Requests Distribution}\label{sec:distributor}
As detailed in Section~\ref{sec:latency_vs_depth}, the workload distribution policy should prevent requests from experiencing cascaded timeout due to improper instance assignments. The distributor enforces SLO-aware distribution through a three-step workflow: 
\begin{enumerate}
\item \textbf{Sub-cluster mapping}: Classify requests by SLO requirements into $R_t$ and $R_l$, then route them to corresponding sub-clusters $G_t$ and $G_l$ established by the placer.
\item \textbf{Instance assignment}: For active instances within the target sub-cluster, evaluate each instance's capability to meet the request's SLO and select the instance with the shortest request queue to achieve load balancing.
\item \textbf{Overflow protection}: Block request $r$ assignment when $L_q^r + L_d^r > \tau_r$ is predicted, where $L_d^r$ is estimated using worst-case instance throughput $F(M,P,\mathcal{B}_I)$, so as to prevent cascaded timeouts caused by optimistic estimation.
\end{enumerate}

This distributor workflow is illustrated in Fig.~\ref{fig:System_overview} using the two-sub-cluster example from Section \ref{sec:deploy_partiton}. For example, request $R^1_a \in R_t$ for model $M_a$ is routed to instance $I^t_a$ in $G_t$. The conservative estimation ensures that $L^r_d$ prediction preserves time margins, preventing the propagation of cascaded timeouts during continuous batching.

\section{Experiments}
\subsection{Experimental Setting}\label{sec:exp_setting}

\textbf{Hardware Environment.} Our experiments utilize two servers: (1)  a server equipped with two Intel Xeon E5-2680 CPUs, 128GB of memory, two NVIDIA GeForce RTX 3090 GPUs (24GB each), and (2) a server equipped with two Intel Xeon Platinum 8468V CPUs, 2048GB of memory, eight NVIDIA HGX H20 GPUs (96GB each). All LLM instances employ vLLM as the inference engine with bfloat16 quantization. Since conducting experiments on a MaaS production cluster is prohibitively expensive, we perform cluster-scale simulation experiments using profiling data from our own servers. The simulated cluster is composed of multiple AWS EC2 p3.16xlarge instances\cite{Amazon_EC2}, each equipped with 8 NVIDIA Tesla V100 GPUs (16GB each).
\begin{table}[htbp]
\centering
\caption{Experimental Trace Configurations}
\label{tab:trace_configs}
\begin{tabular}{c|c c c}
\hline
\textbf{Trace}  & \textbf{Decode Length} & \textbf{SLO Factor} & \multirow{2}{*}{\textbf{Proportion}} \\

\textbf{No.}  & \textbf{($S_r$ tokens)} & ($\theta_r$) &  \\
\hline
\hline
\multirow{2}{*}{\textbf{1}} 
 & \multirow{2}{*}{300-1000}  & \multirow{2}{*}{0.8-1.5} & \multirow{2}{*}{$-$} \\
 &  &  & \\
\hline
\multirow{2}{*}{\textbf{2}} 
 & \multirow{2}{*}{300-500} & 0.8-1.0 & \multirow{2}{*}{$-$} \\
 &  & 1.2-1.5 &  \\
\hline
\multirow{2}{*}{\textbf{3}} 
 & 300-500 & \multirow{2}{*}{0.8-1.2} & \multirow{2}{*}{$-$} \\
 & 600-1000 &  &  \\
\hline
\multirow{2}{*}{\textbf{4}} 
  & 300-500 & 0.8-1.0 & 50\% \\
 & 600-1000 & 1.2-1.5 & 50\% \\
\hline
\multirow{2}{*}{\textbf{5}} 
 & \multirow{2}{*}{300-500} & 0.8-1.0 & 34\% \\
 &  & 1.2-1.5 & 66\% \\
\hline
\multirow{2}{*}{\textbf{6}} 
 & \multirow{2}{*}{300-500} & 0.8-1.0 & 66\% \\
 &  & 1.2-1.5 & 34\% \\
\hline
\end{tabular}
\end{table}


\begin{figure*}[htbp] 
    \centering
    \includegraphics[width=0.95\linewidth]{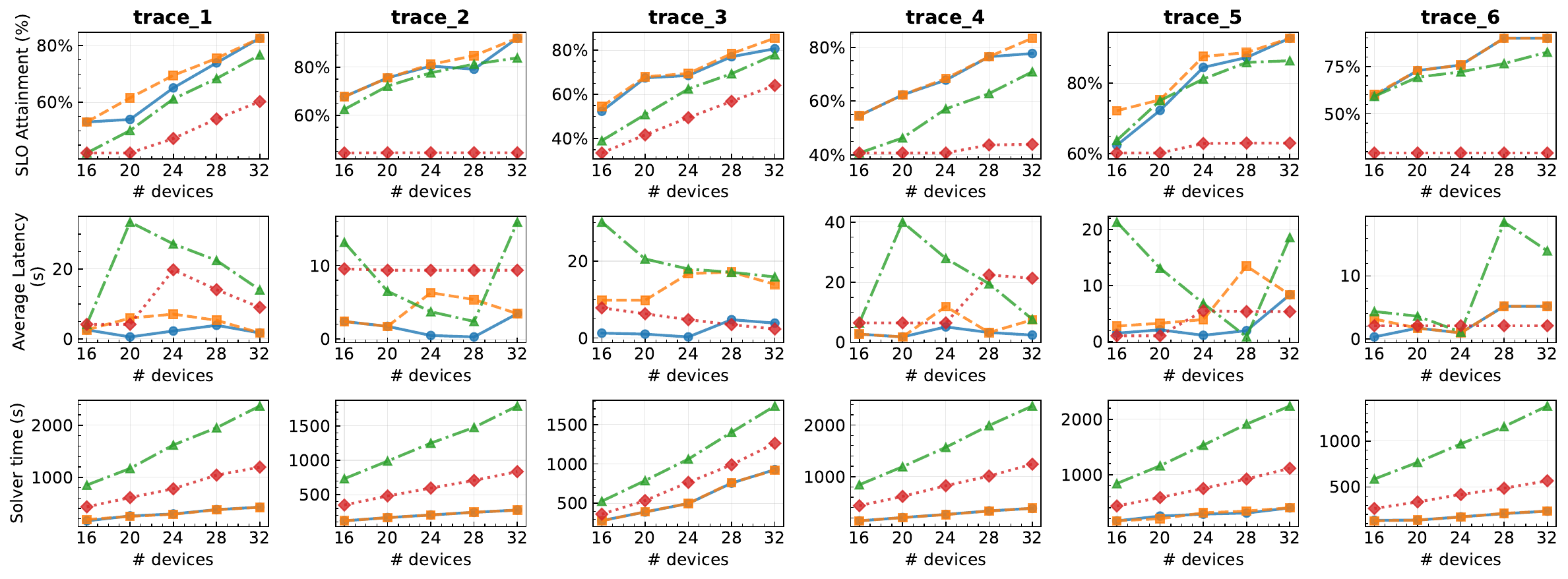}
    \includegraphics[width=0.95\linewidth]{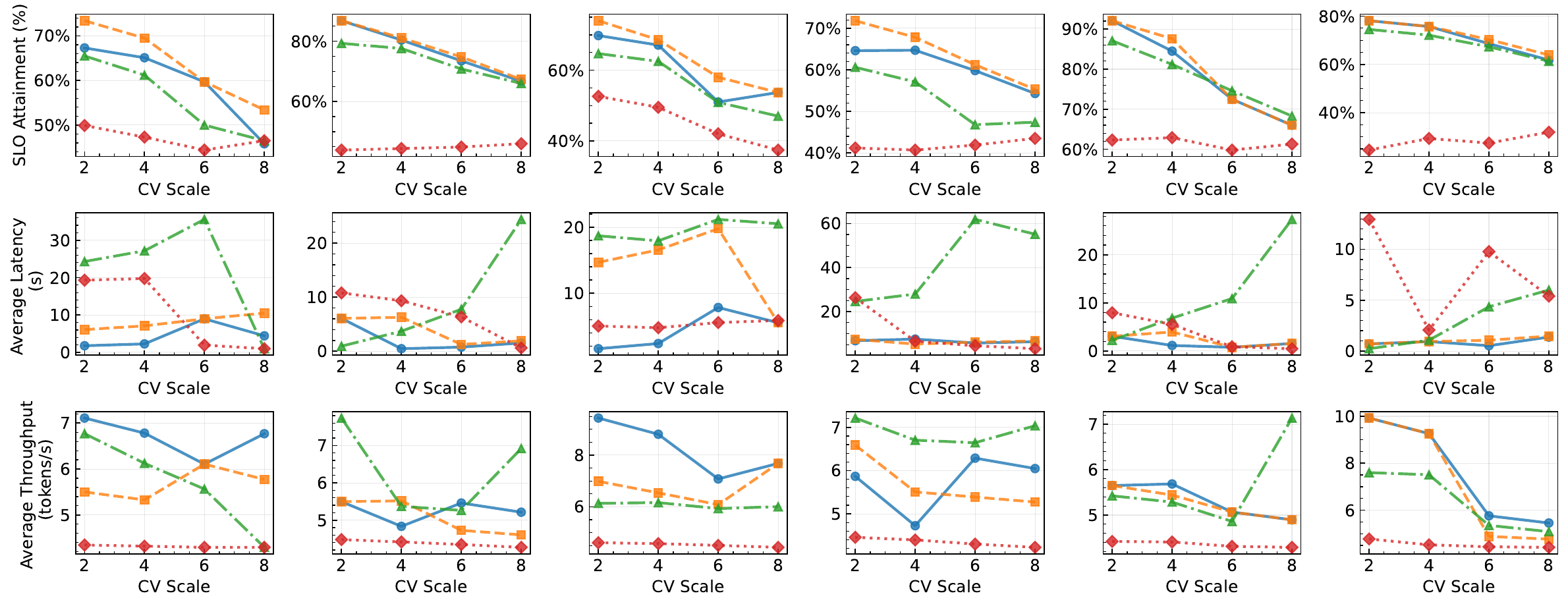}
    \includegraphics[width=0.95\linewidth]{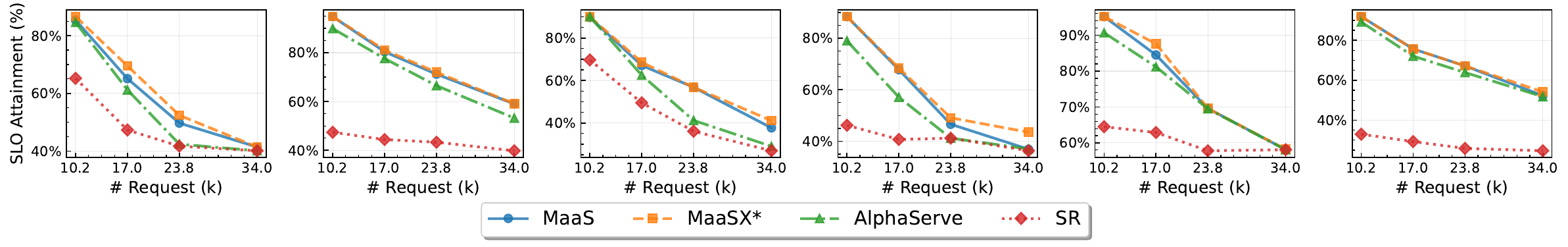}
    \caption{Experimental results in three different scenarios.} \label{fig:exp_3_scenarios}
\end{figure*}

\textbf{Workload Traces.} We generate realistic inference workloads for three Transformer-based LLMs by leveraging established traffic sampling protocols\cite{Alpaserve,Clockwork,InferLine}. Request arrivals within each time window follow a gamma process parameterized by the sampling rate and the coefficient of variance ($CV$). To simulate diverse SLO requirements and ensure fairness despite unpredictable output lengths of LLMs, we randomly assign a decoding length $S_r$ and an SLO factor $\theta_r$ to each request $r$. For the sake of fairness, the performance profiles are derived from systematic measurements on our servers and aligned with the profile data published by AlpaServe\cite{Alpaserve}. As shown in Table~\ref{tab:trace_configs}, we construct six representative traces that simulate different MaaS scenarios, including varying SLO strictness, diverse contextual complexity, and non-uniform request proportions. 

\textbf{Setup.} By default, the cluster comprises 24 GPU devices. The workload sampler is configured with $CV = 2$ and a sampling rate tuned to capture $17,000$ request arrivals within each $3600$-second window. To evaluate MaaSO under diverse conditions, we independently vary one of these variables (i.e., cluster size, $CV\_$scale, or sampling rate) to assess its performance across different cluster scales, request burst patterns, and overall workload pressures. MaaSO prioritizes response latency optimization through heterogeneous instance orchestration, where the serving score is configured as $\alpha = 4$ and $\beta = 0.3$.

\textbf{Baseline Methods.} Since there are no existing works focusing on heterogeneous instance orchestration, we select \textit{AlpaServe} \cite{Alpaserve} and \textit{Selective Replication (SR)} as baselines. As discussed in Section \ref{sec:related}, AlpaServe is designed to provide globally optimal homogeneous placement solution based on SLO requirements and performs load-balanced request allocation. In contrast, \textit{Selective Replication (SR)} implements DP-instance placement without parallelism policies search, which mimics the policy of a wide range of existing serving systems\cite{SR1,SR2,DARKBIRD}. To ensure comparability with our design, both these two algorithms are extended with inference batch size searching and the same search space pruning strategies as MaaSO. We also use \textit{MaaSO*} as an ablation baseline. It sets $\alpha=10$ in Eq.~\eqref{eq:score_simulation}, prioritizing SLO satisfaction over throughput and latency optimization to identify optimal SLO attainment through heterogeneous instance configuration. 

\textbf{Evaluation Metrics.} We evaluate system performance using four key metrics: (1) \textit {SLO Attainment}: the ratio of requests with satisfied SLO requirements; (2) \textit{Average Decoding Throughput (tokens/s)}: the average rate at which the system decodes tokens; (3) \textit{Average Response Latency (s)}: the average time to deliver the first token of a response; (4) \textit{Solver overhead (s)}: the time consumed by the placer to configure instances for a given request set.



\textbf{Simulator Implementation.} We implement a discrete-event simulator that extends AlpaServe's framework to support the autoregressive decoding characteristics of LLMs. Due to the computational complexity of token-by-token simulation, we design a virtual-slot-based approach, where the number of virtual slots is determined by the inference batch size $\mathcal{B}_I$. The simulator employs a map-reduce strategy for request processing: the distributor routes requests to appropriate sub-clusters and model instances based on SLO requirements. Then, the map function assigns requests to available slots, and the reduce function advances instance clocks when slots are unavailable and re-attempts mapping. If no slots remain available, the reduce operation continues until the request is either accommodated or rejected due to insufficient remaining time for completion.

\subsection{Results and Analysis} \label{sec:exp_analysis}
As shown in Fig.~\ref{fig:exp_3_scenarios}, we evaluate MaaSO against baseline methods across six traces in three experimental scenarios: varying cluster scales (Rows 1-3), different request burstiness (Rows 4-7), and the total number of requests arrived within serving duration (Last Row).

\textbf{SLO Attainment}: Results in the first row demonstrate that MaaSO* and MaaSO achieve significant advantages in SLO satisfaction across most traces, with Traces 3 and 4 showing the most pronounced improvements. This indicates that homogeneous approaches are insufficient for MaaS scenarios involving diverse decoding lengths, as MaaSO satisfies more request SLOs under resource constraints. However, Traces 5 and 6 (generated using a piecewise request distribution function) favor homogeneous strategies for satisfying majority request SLOs. Additionally, among all traces, Trace 2 exhibits the least SLO diversity, yet MaaSO still achieves comparable or superior performance. Experiments in rows 4 and 7 reveal that homogeneous approaches exhibit rapid performance degradation under higher workload pressure and request burstness, while MaaSO maintains stable SLO attainment compared to these baselines.



\textbf{Response Latency \& Decoding Throughput}: Rows 2 and 5 reveal the average response latency for user requests. Response times exceeding ten seconds are typically unacceptable to users, yet baseline methods lack optimization for this metric in workload distribution and instance deployment, resulting in highly unpredictable request queuing times. MaaSO achieves both higher SLO attainment and globally lowest response latency through its heterogeneous instance orchestration mechanism. Compared to the ablation baseline MaaSO*, MaaSO maintains comparable SLO satisfaction while providing more stable response latency guarantees, even under high request burstness (row 5). Since serving score in MaaSO is configured as $\beta=0.3$, it does not achieved the globally optimal decoding throughput in row 6. 

\textbf{Solver Overhead}: In row 3, we compare the solver's overhead as cluster scale increases. Although we equip the baselines with the same search space optimization strategies, these homogeneous approaches still require substantial decision time when the cluster scale expands. MaaSO's sub-cluster partition paradigm effectively filters out more invalid configurations. Compared to AlpaServe, SR reduces the overhead of configuration selection by eliminating the parallelism strategy decisions. However, solver overhead exceeding 1000s on 32-GPU clusters is significantly slower than model instance startup times (e.g., 10 minutes). In contrast, MaaSO demonstrates substantially lower computational overhead compared to baseline methods. The ablation baseline MaaSO* has the same algorithmic complexity as MaaSO, highlighting the high configurability of our proposed method.


\textbf{Limitations Analysis:} Through extensive experiments, we identified that MaaSO, as a heterogeneous instance orchestrator, exhibits performance degradation in two specific cases. Firstly, when the requests SLO diversity reduced (as in Trace 5, Row 4), the performance benefits of heterogeneous instances deployed through complementary strategies diminish. Secondly, when massive requests arrive uniformly, the performance convergence occurs. As shown in the last row experiments, when the total requests reach $34k$, all approaches (including the DP-based approach SR) achieve similar SLO attainment. This represents the performance convergence mentioned in Fig.~\ref{fig:Instance_interference}, where TP-instances no longer gain benefits in serving performance and instance service slots remain fully saturated. However, with SLO attainment below 50\%, such cluster size is unreasonable for handling inference requests of this magnitude.


\section{Conclusion}
This paper presents MaaSO, the first MaaS orchestrator that enables heterogeneous instance orchestration for serving LLM inference tasks with diverse SLO requirements. Composed of a profiler, a placer, and a distributor, MaaSO's modular design ensures efficient request distribution while supporting complex heterogeneous instance configurations. MaaSO addresses three core challenges: performance modeling of parallel instances under varying workload level, the optimization of heterogeneous instance configuration, and SLO-aware request distribution across heterogeneous instances while preventing cascaded timeouts in continuous batching. 




\bibliographystyle{ieeetr}
\bibliography{ref}

\end{document}